\documentclass{article}
\usepackage{xcolor, graphicx}
\usepackage{amsmath}
\usepackage{caption,subcaption}
\usepackage[figuresright]{rotating}
\usepackage{longtable}
\usepackage{geometry}
\usepackage{amsfonts}

\usepackage[toc,page]{appendix}

\renewcommand{\theequation}{\arabic{section}.\arabic{equation}}

\title{ Thermodynamic limit and twisted boundary energy of the XXZ spin chain with antiperiodic boundary condition}
\author{Zhirong Xin${}^{a}$, Yi Qiao${}^{a}$,  Kun Hao${}^{a,b}$, Junpeng Cao${}^{c,d,e}$, Wen-Li Yang${}^{a,b,f}\footnote{Corresponding author:
wlyang@nwu.edu.cn}$, \\ Kangjie Shi${}^{a,b}$ and Yupeng Wang${}^{c,d,e}\footnote{Corresponding author: yupeng@iphy.ac.cn}$}

\begin{document}
\date{}
\maketitle
\begin{center}
     ${}^a$Institute of Modern Physics, Northwest University,
     Xian 710069, China\\
     ${}^b$Shaanxi Key Laboratory for Theoretical Physics Frontiers,  Xian 710069, China\\
     ${}^c$Beijing National Laboratory for Condensed Matter
           Physics, Institute of Physics, Chinese Academy of Sciences, Beijing
           100190, China\\
     ${}^d$School of Physical Sciences, University of Chinese Academy of Sciences, Beijing, China\\
     ${}^e$Collaborative Innovation Center of Quantum Matter, Beijing,
     China\\
     ${}^f$School of Physics, Northwest University, Xi'an 710069, China
\end{center}

\begin{abstract}
We investigate the thermodynamic limit of the inhomogeneous $T-Q$ relation of the antiferromagnetic XXZ spin chain with antiperiodic boundary condition. It is shown that the contribution of the inhomogeneous term for the ground state can be neglected when the system-size $N$ tends to infinity, which enables us to reduce the inhomogeneous Bethe ansatz equations (BAEs) to the homogeneous ones. Then the quantum numbers at the ground states are obtained, by which the system with arbitrary size can be studied. We also calculate the twisted boundary energy of the system.

\vspace{0.5truecm}
\end{abstract}

\section{Introduction}

The XXZ spin chain with the antiperiodic boundary condition (or the twisted boundary condition) is a very interesting quantum system \cite{Yung1995, Batchelor1995, Niekamp2009, Niccoli2013}. By using the Jordan-Wigner transformation, the model can describe a p-wave Josephson junction embedded in a spinless Luttinger liquid \cite{Fazio1995, Winkelholz1996, Caux2002}. Although there exists a twisted bound at the boundary which breaks the usual $U(1)$-symmetry  of the bulk  system (or the closed chain case) \cite{Yupeng2015}, it can be proved that the system is still integrable. By using the off-diagonal Bethe ansatz (ODBA) method \cite{ODBAPhys.Rev.Lett.111, ODBANucl.Phys.B875, ODBANucl.Phys.B877}, the exact solution of the model was obtained \cite{ODBAPhys.Rev.Lett.111}, which is described by an inhomogeneous $T-Q$ relation (c.f. the ordinary homogeneous $T-Q$ one \cite{Bax82, Kor93}).  Such an inhomogeneous $T-Q$ relation has played  a universal role to describe the eigenvalue of the transfer matrix for quantum integrable systems \cite{Yupeng2015}.
However, due to the fact that Bethe roots should satisfy the inhomogeneous Bethe ansatz equations (BAEs), it is hard to study the thermodynamic properties \cite{Takahashi} of the corresponding systems \cite{Nep13, Jia13, Nep14}.

Based on an intelligent trick, the  thermodynamic limit of  the spin-$\frac{1}{2}$ XXZ chain with the generic off-diagonal boundary terms in the gapless region (i.e., the anisotropy parameter $\eta$ in (\ref{Hamiltoniant}) below being an imaginary number)  was succeeded in obtaining \cite{Li14}.  The most important observation in the paper is that the contribution of the inhomogeneous term for the ground state, in the gapless region, can be neglected when the system-size $N$ tends to infinity.  Such a fact has been confirmed recently by the studies of other integrable models \cite{Wen17, Wen17-1, Pir18, Takahas11hi} whose eigenvalue of the transfer matrix is given in terms of the inhomogeneous $T-Q$ relation.

In this paper, we propose a method to study the thermodynamic limit of the XXZ spin chain with the twisted boundary condition at the antiferromagnetic region (i.e., $\eta$ being a real number). We first study the contribution of the inhomogeneous term with finite system-size $N$. We find that the contribution of the inhomogeneous term in the associated $T-Q$ relation to the ground state energy can be neglected when the system-size $N$ tends to infinity. Because we consider the massive region of the system, the ground state energy with even $N$ and that with odd $N$ are different. The value of energy difference is proportional to the energy of one bond.
We also check our results by using the density matrix renormalization group (DMRG) method \cite{dmrg White 1993,dmrg 2005},  which leads to that the numerical results and the analytic one are consistent with each other very well. As a consequence, we obtain the  twisted boundary energy of the model.

The paper is organized as follows. In the next section, the model and the associated ODBA solutions are introduced.
In section 3, we study the finite-size effects of contribution of the inhomogeneous term in the $T-Q$ relation for the ground state.
The thermodynamic limit of the XXZ spin chain with antiperiodic and with periodic boundary conditions are discussed in section 4 and section 5, respectively.
The twisted boundary energy is given in Section 6. Section 7 is the concluding remarks and discussions. Some supporting detailed calculations are given in
Appendices A$\&$B.

\section{The model and its ODBA solution}

\setcounter{equation}{0}

The spin-$\frac{1}{2}$ XXZ quantum  chain is described by the Hamiltonian
\begin{equation}\label{Hamiltoniant}
H=\sum^N_{j=1} \left[  \sigma^x_j \sigma^x_{j+1} + \sigma^y_j \sigma^y_{j+1} +\cosh\eta \sigma^z_j \sigma^z_{j+1} \right],
\end{equation}
where the antiperiodic boundary condition reads $\sigma^{\alpha}_{N+1} =\sigma^x_1 \sigma^{\alpha}_1 \sigma^x_1~(\alpha=x,y,z)$, and $\sigma^{\alpha}_j$ is the Pauli matrix.
For such a topological boundary condition (c.f. the periodic boundary condition), the spin on the $N$th site couples with that on the first site after rotating by an angle $\pi$ along the $x$-direction (a kink on the $(N,1)$ bond) and the system forms a m\"obius strip in the spin space. This kink could be shifted to the $(j,j+1)$ bond with the spectrum of the Hamiltonian unchanged
\begin{equation}
\tilde{H}_j=U^x_j H U^x_j, \qquad U^x_j=\prod^j_{l=1}\sigma^x_l.
\end{equation}
Due to the fact $[H, U_N^x]=0$, the model possesses a global $Z_2$ invariance. Note that the braiding occurs in the quantum space rather than in the real space.

The integrability of the model (\ref{Hamiltoniant}) is associated with the well-known six-vertex $R$-matrix
\begin{eqnarray}
  R_{0,j}(u) &=& \frac{1}{2} \left[ \frac{\sinh(u+\eta)}{\sinh \eta} (1+\sigma^z_j \sigma^z_0) +\frac{\sinh u}{\sinh \eta} (1- \sigma^z_j \sigma^z_0)   \right] \nonumber \\
   && + \frac{1}{2} (\sigma^x_j \sigma^x_0 +\sigma^y_j \sigma^y_0),
\end{eqnarray}
where $u$ is the spectral parameter and $\eta$ is the crossing parameter (or the anisotropy parameter).
The $R$-matrix satisfies the Yang-Baxter equation
\begin{equation}
  R_{1,2}(u-v) R_{1,3}(u) R_{2,3}(v)=R_{2,3}(v) R_{1,3}(u) R_{1,2}(u-v),
\end{equation}
and possesses the properties:
\begin{eqnarray}
  \text {Initial condition:} && R_{1,2}(0)=P_{1,2}, \\
  \text {Unitarity:} &&  R_{1,2}(u)R_{2,1}(-u)=- \frac{\sinh(u+\eta)\sinh(u-\eta)}{\sinh^2 \eta} \times \text{id} , \\
  \text {Crossing ~relation:} && R_{1,2}(u)=-\sigma^y_1 R^{t_1}_{1,2}(-u-\eta)\sigma^y_1 , \\
  \text {$Z_2$-symmetry:} && \sigma^\alpha_1 \sigma^\alpha_2 R_{1,2}(u)= R_{1,2}(u)  \sigma^\alpha_1 \sigma^\alpha_2 ,~~~\text{for} ~ \alpha=x,~y,~z, \label{Z2symmetry}
\end{eqnarray}
where $P_{1,2}$ is the permutation operator, and $t_i$ denotes the transposition in the $i$th space.
Here and below we adopt the standard notations: for any matrix $A\in {\rm End}(C^2)$, $A_i$ is an operator embedded in the tensor space $C^2\otimes C^2\otimes\cdots$, which acts as $A$ on the $i$-th space and as identity on the other factor spaces;
$R_{i,j}(u)$ is an operator of R-matrix embedded in the tensor space, which acts as identity on the factor spaces except for the $i$-th and $j$-th ones.

The associated monodromy matrix is given as
\begin{equation}
  T_0(u)=\sigma_0^x R_{0,N}(u-\theta_N) \cdots R_{0,1}(u-\theta_1)=\left(
                                                                     \begin{array}{cc}
                                                                       C(u) & D(u)\\
                                                                        A(u)& B(u)\\
                                                                     \end{array}
                                                                   \right).
\end{equation}
Because of the $Z_2$-symmetry (\ref{Z2symmetry}), the following relation holds
\begin{equation}
  R_{0, \bar{0} }(u-v) T_0(u)T_{\bar{0}}(v)=T_{\bar{0}}(v) T_0(u) R_{0, \bar{0} }(u-v),
\end{equation}
which directly gives rise to the fact that
\begin{equation}
  [t(u),t(v)]=0,
\end{equation}
where the transfer matrix $t(u)$ is defined as
\begin{equation}\label{traM}
  t(u)=tr_0 T_0(u)=B(u)+C(u) .
\end{equation}
The first order derivative of the logarithm of the transfer matrix gives  Hamiltonian (\ref{Hamiltoniant})
\begin{equation}
  H=2 \sinh \eta \frac{\partial \ln t(u)}{\partial u} \mid_{u=0, \{ \theta_j=0 \}} - N \cosh \eta,
\end{equation}
with the antiperiodic boundary condition. This ensures the integrability of the model.

By means of the off-diagonal Bethe ansatz method, the eigenvalues $\Lambda(u)$ of the transfer matrix $t(u)$ is given by  the inhomogeneous $T-Q$ relation \cite{Yupeng2015}
\begin{equation}\label{inhTQ}
  \Lambda(u) = e^u a(u) \frac{Q(u-\eta)}{Q(u)} -  e^{-u-\eta} d(u) \frac{Q(u+\eta)}{Q(u)} - c(u)a(u) d(u) \frac{1}{Q(u)},
\end{equation}
where $Q(u)$ is a trigonometric polynomial of the type
\begin{equation}\label{inhTQ11}
  Q(u)=\prod^N_{j=1} \frac{\sinh(u-\lambda_j)}{\sinh \eta},
\end{equation}
and
\begin{eqnarray}
d(u)=a(u-\eta)=\prod^N_{j=1} \frac{\sinh(u-\theta_j)}{\sinh \eta},\quad c(u)=e^{u-N\eta+\sum^N_{l=1}(\theta_l-\lambda_l) } -e^{-u-\eta -\sum^N_{l=1}(\theta_l-\lambda_l) }.
\end{eqnarray}
The $N$ parameters $\{ \lambda_j \}$ in Eq. (\ref{inhTQ11}) should satisfy the associated BAEs
\begin{eqnarray}
  && e^{\lambda_j} a(\lambda_j) Q(\lambda_j-\eta) -  e^{-\lambda_j-\eta} d(\lambda_j) Q(\lambda_j+\eta) - c(\lambda_j)a(\lambda_j) d(\lambda_j) = 0, \nonumber\\[5pt]
  &&~~~~~~j=1, \cdots ,N.\label{Inhomogeneous BAE}
\end{eqnarray}

The eigenvalue of the Hamiltonian (\ref{Hamiltoniant}) is then expressed in terms of the associated Bethe roots as
\begin{eqnarray} \label{ene1119}
E&=&2 \sinh \eta \frac{\partial \ln \Lambda(u)}{\partial u} \mid_{u=0, \{ \theta_j=0 \}} - N \cosh \eta \nonumber \\
&=&-2 \sinh \eta \sum^N_{j=1} [\coth(\lambda_j +\eta) -\coth (\lambda_j)] +N \cosh \eta +2 \sinh \eta,
\end{eqnarray}
where the Bethe roots $\{ \lambda_j \}$ satisfy the inhomogeneous Bethe ansatz equations (BAEs) (\ref{Inhomogeneous BAE}),
and the numerical simulation implies that the inhomogeneous BAEs (\ref{Inhomogeneous BAE}) indeed give the correct and
complete spectrum of the model \cite{Yupeng2015}.

\section{Finite-size effects}
\setcounter{equation}{0}

In this paper, we consider the massive region, namely, with a real $\eta$.
In order to study the contribution of the inhomogeneous term (i.e., the last term in Eq. (\ref{inhTQ})) to the ground state energy, we first introduce  a homogeneous $T-Q$ relation as
\begin{equation}\label{homTQ}
  \Lambda_{hom}(u)= e^u a(u) \frac{Q_1(u-\eta)}{Q_1(u)} -  e^{-u-\eta} d(u) \frac{Q_1(u+\eta)}{Q_1(u)},
\end{equation}
where
\begin{equation}
  Q_1(u)=\prod^M_{j=1} \sinh(u-\lambda_j).
\end{equation}
It should be remarked that the number of Bethe roots in Eq. (\ref{homTQ}) is reduced to $M$ ($ M\leq N $).
The singularity analysis of the $T-Q$ relation (\ref{homTQ}) gives rise to  the homogeneous BAEs (c.f., (\ref{Inhomogeneous BAE}))
\begin{equation}\label{BAEt1}
  e^{i \eta x_j } \frac{\sin^N\frac{\eta}{2}(x_j-i)}{ \sin^N\frac{\eta}{2}(x_j+i) } =\prod^M_{k=1} \frac{\sin\frac{\eta}{2}(x_j-x_k-2i) }{ \sin\frac{\eta}{2}(x_j-x_k+2i) },
  \quad j=1,\ldots,M,
\end{equation}
where the transformation $\lambda_j= \frac{\eta}{2} ( i x_j -1)$ is used.

Taking the logarithm of Eq. (\ref{BAEt1}), we have
\begin{equation}\label{logBAEt}
  \eta x_j + N \theta_1(x_j)=2 \pi I_j +\sum^M_{k=1} \theta_2(x_j - x_k), \quad j=1,\ldots,M,
\end{equation}
where
\begin{equation}\label{theta}
  \theta_m(x)= 2\arctan \frac{\tan \frac{\eta x}{2}}{\tanh \frac{\eta m}{2}} +2\pi \left[\frac{\eta x +\pi}{2 \pi} \right].
\end{equation}
Here the notation $[~ ]$ represents the Gauss Mark, and the quantum number $\{ I_j \}$ are certain integers (half odd integers) for $N-M$ even ($N-M$ odd).
Corresponding to Eq. (\ref{ene1119}), we define
\begin{eqnarray} \label{eigenvaluet1}
E_{hom}&=&2 \sinh \eta \frac{\partial \ln \Lambda_{hom}(u)}{\partial u} \mid_{u=0} - N \cosh \eta \nonumber \\
&=& -4 \sinh \eta \sum^M_{j=1} \frac{\sinh \eta}{\cosh \eta - \cos(\eta x_j)}  + N \cosh \eta +2 \sinh\eta.
\end{eqnarray}

Now, we define the contribution of the inhomogeneous term to the ground state energy as
\begin{equation}\label{inhE}
  E^g_{inh}\equiv E^{g}_{hom} - E^{g},
\end{equation}
where $E^{g}$ is the ground state energy of the Hamiltonian (\ref{Hamiltoniant}) and $E^{g}_{hom}$ is the minimal energy calculated by Eqs.(\ref{logBAEt}) (or the reduced BAEs (\ref{BAEt1})) and (\ref{eigenvaluet1}).
We remark  that the ground state energy $E^{g}$ for some small system-size $N$ can be calculated by BAEs (\ref{Inhomogeneous BAE}) and relation (\ref{ene1119}) or by direct diagonalization of the Hamiltonian ($N$ up to 24), while for some large system-size $N$ (up to 100) we use the DMRG method\footnote{We have used the infinite chain DMRG algorithm \cite{dmrg White 1993,dmrg 2005}, starting with 14 sites, the number of reserved states $m=2^7$, and the truncation error is $10^{-9}$. In the small $N$ (less than 24) case, the DMRG results are in pretty good agreement with the direct diagnalization results (the relative errors in ground state energies is $10^{-10}$ ), which is enough for our numerical analyses.} \cite{dmrg White 1993,dmrg 2005} to calculate the ground state energy $E^{g}$ instead.

\begin{figure}[!htp]
    \begin{minipage}[t]{0.5\linewidth}
    \centering
    \includegraphics[height=4.5cm,width=8cm]{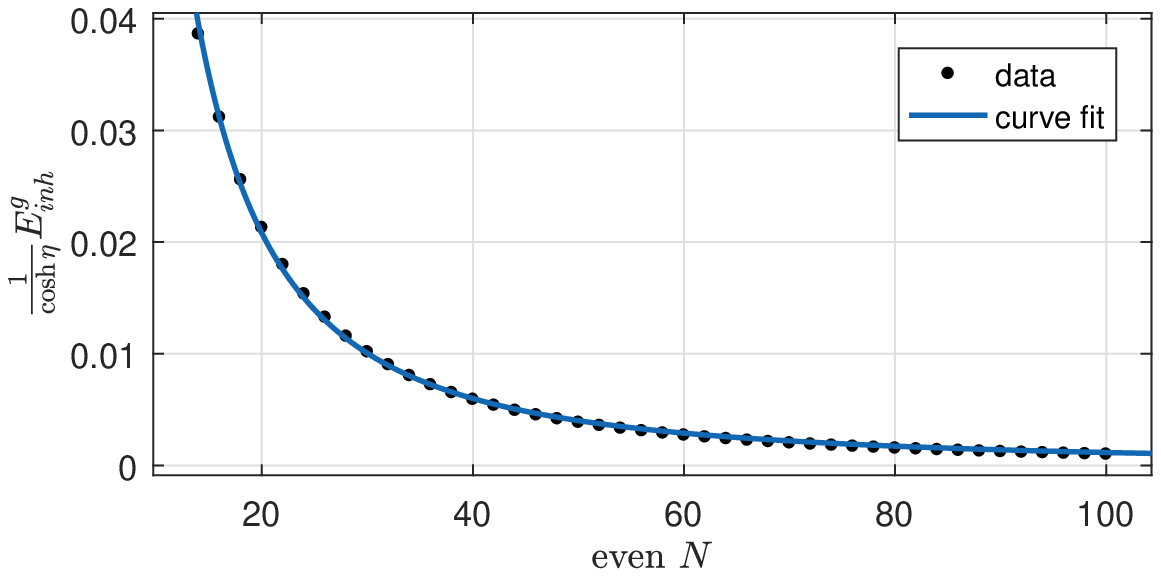}
    \caption*{(a)}
    \end{minipage}
    \begin{minipage}[t]{0.5\linewidth}
    \centering
    \includegraphics[height=4.5cm,width=8cm]{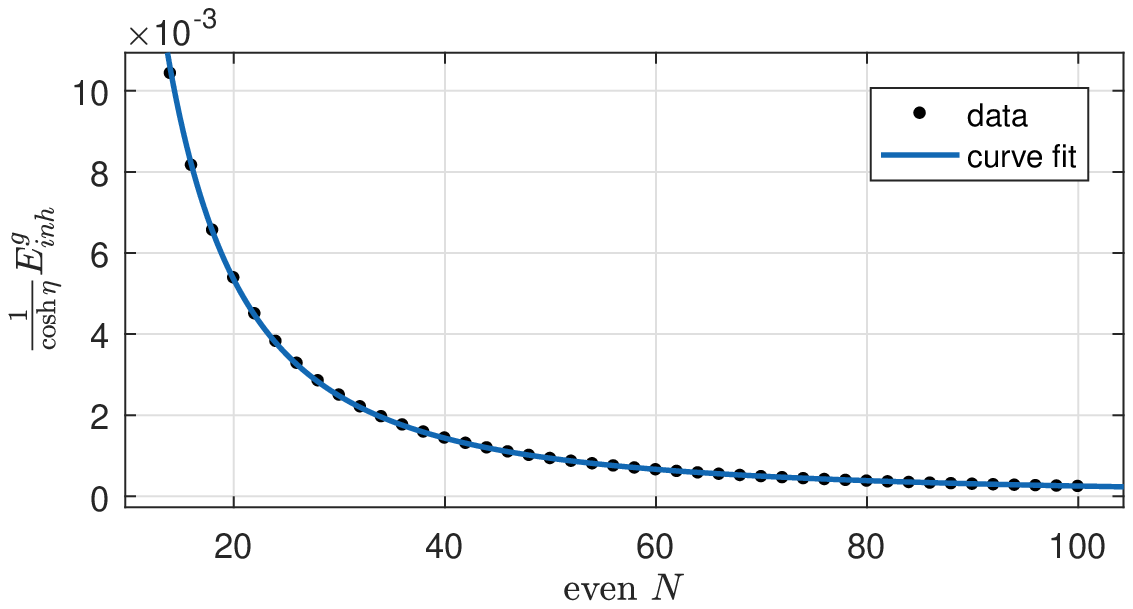}
    \caption*{(b)}
    \end{minipage}
    \caption{The contribution of the inhomogeneous term to the ground state energy $\frac{1}{\cosh \eta} E^g_{inh}$ versus the even system-size $N$.
The data can be fitted as $ \frac{1}{\cosh \eta} E^g_{inh}(N)=a_1 N^{b_1}$.
Here (a) $\eta=2$, $a_1=4.525$ and $b_1=-1.797$; (b) $\eta=3$, $a_1=1.58$ and $b_1=-1.899$. Due to the fact $b_1< 0$, the contribution of the inhomogeneous term tends to zero when the $N \rightarrow \infty$.}\label{fig-FC-even}
\end{figure}

Because we consider the massive region, the thermodynamic limit of the system with even $N$ and that with odd $N$ are different.
We first consider the even $N$ case. It is found that for this case the minimal energy $E^{g}_{hom}$ calculated by Eqs.(\ref{eigenvaluet1}) and (\ref{logBAEt})
is given by  $M=\frac{N}{2}$ and all the Bethe roots in Eq. (\ref{logBAEt}) being real and   the corresponding quantum numbers being
\begin{equation}\label{qnumber-t-g-even}
  I_j=-\frac{M}{2}+1, -\frac{M}{2}+2, \cdots , \frac{M}{2}.
\end{equation}
Substituting the values of Bethe roots into Eq.(\ref{eigenvaluet1}), we obtain the value of $E^{g}_{hom}$. From Eq. (\ref{inhE}), the contribution of the inhomogeneous term
can be calculated and the results are shown in Fig. \ref{fig-FC-even}. From the fitting, we find that $E^g_{inh}$ and $N$ satisfy the power law\footnote{The numerical simulations imply that $b_1<-1$ for all real $\eta$ and become smaller with increasing $\eta$, which means that the contribution of the inhomogeneous terms tend to zero more faster for a larger $\eta$. Moreover, $b_1$ as a function of $\eta$ is too complicated to be determined. It is interesting to note that the homogeneous Bethe system can replace the inhomogeneous one for the purpose of computing finite-size corrections up to order of unity.}
\begin{equation}
\frac{ 1}{\cosh \eta} E^g_{inh}(N)=a_1 N^{b_1}.\label{b1}
\end{equation}
Due to the fact that $b_1<0$, the value of $E^g_{inh}$ tends to zero when the system-size $N$ tends to infinity, which means that the contribution of the inhomogeneous term for the ground state can be neglected in the thermodynamic limit.

\begin{figure}[!htp]
    \begin{minipage}[t]{0.5\linewidth}
    \centering
    \includegraphics[height=4.5cm,width=8cm]{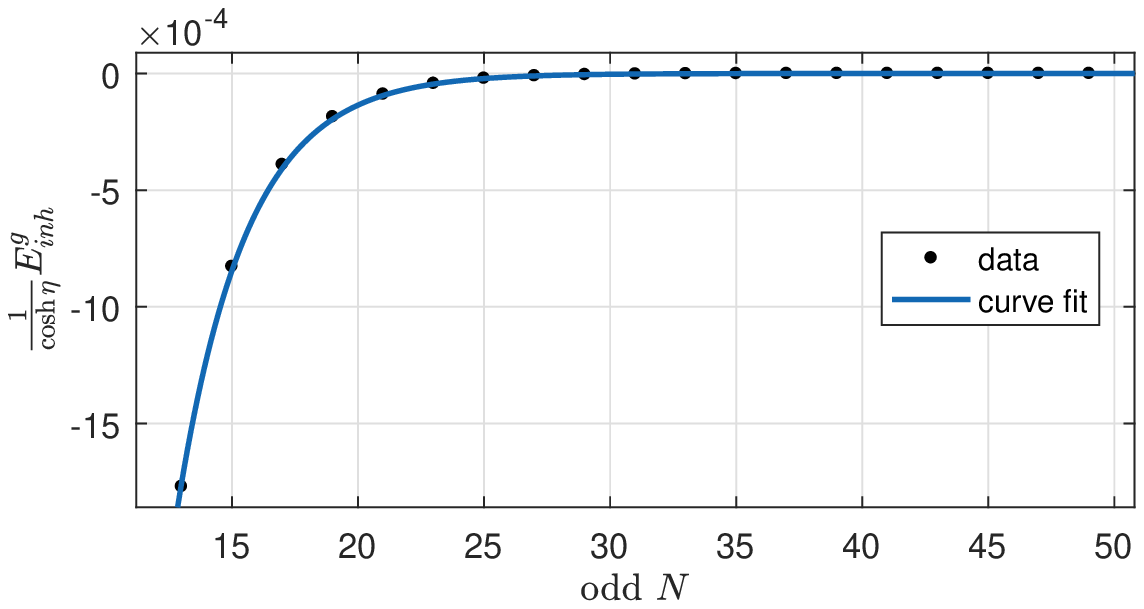}
    \caption*{(a)}
    \end{minipage}
    \begin{minipage}[t]{0.5\linewidth}
    \centering
    \includegraphics[height=4.5cm,width=8cm]{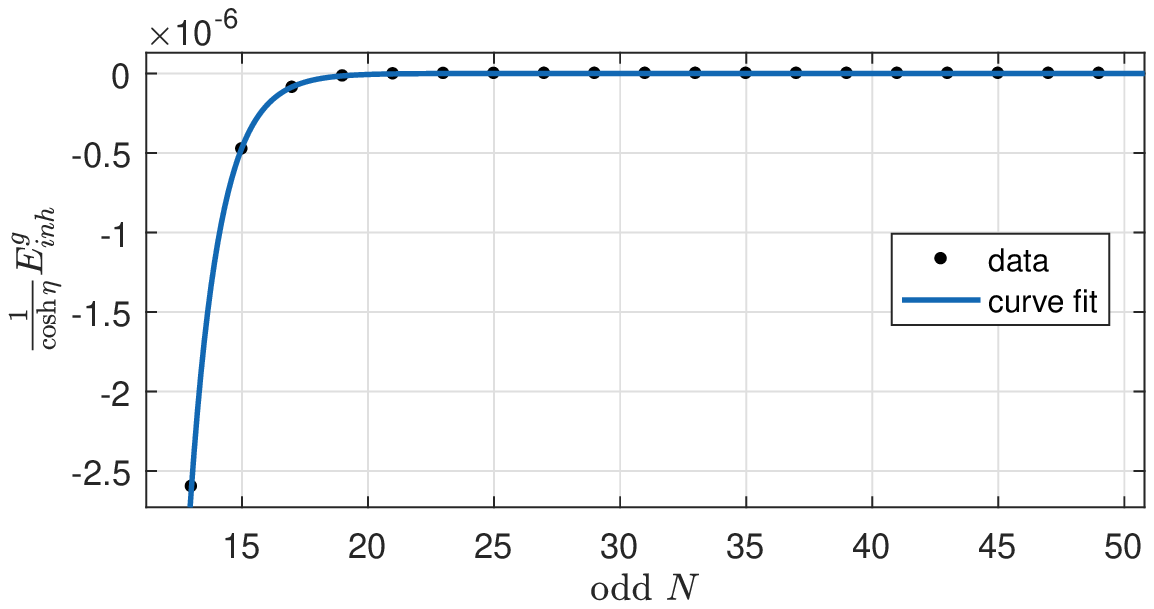}
    \caption*{(b)}
    \end{minipage}
    \caption{The contribution of the inhomogeneous term to the ground state energy $\frac{1}{\cosh \eta}E^g_{inh}$ versus the odd system-size $N$.
The data can be fitted as $ \frac{1}{\cosh \eta}E^g_{inh}(N)=a_2 e^{b_2N}$. Here (a) $\eta=2$, $a_2=-0.2042$ and $b_2=-0.3658$; (b) $\eta=3$, $a_2=-0.1828$ and $b_2=-0.8585$.
Due to the fact $b_2< 0$, when the $N$ tends to infinity, the contribution of the inhomogeneous term tends to zero.}\label{fig-FC-odd}
\end{figure}

For the odd $N$, we consider the case $M=\frac{N+1}{2}$ in which all the Bethe roots are real.
The $E^{g}_{hom}$ can be calculated by Eq. (\ref{eigenvaluet1}) where the Bethe roots in Eq. (\ref{logBAEt}) are completely determined by the quantum number
\begin{equation}\label{qnumber-t-g-odd}
  I_j=-\frac{M-1}{2}, -\frac{M-1}{2}+1, \cdots , \frac{M-1}{2}.
\end{equation}
The contributions of the inhomogeneous term are shown in Fig. \ref{fig-FC-odd}.
From the fitting, we find that $E^g_{inh}$ and $N$ satisfy the exponential law
\begin{equation}
\frac{ 1}{\cosh \eta} E^g_{inh}(N)=a_2 e^{b_2N}.
\end{equation}
Again, due to the fact that $b_2<0$, the value of $E^g_{inh}$ tends to zero when $N \rightarrow \infty$. Therefore, the contribution of the inhomogeneous term at the ground state can be neglected in the thermodynamic limit.

Through above finite-size scaling analysis, we conclude that the contribution of the inhomogeneous term for the ground state energy can be neglected when $N\rightarrow\infty$ (The similar results have also been obtained \cite{Takahas11hi}). Moreover, from Figs. \ref{fig-FC-even} and \ref{fig-FC-odd}, we also find that $E^g_{inh}>0$ for the even $N$ case and $E^g_{inh}<0$ for the odd $N$ case. This means that $E_{hom}$ is larger than the actual value for the even $N$ case while $E_{hom}$ is smaller than the actual value for the odd $N$ case.
Therefore, the reduced BAEs (\ref{BAEt1}) and Eq. (\ref{eigenvaluet1}) can be used to calculate the ground state energy of the system (\ref{Hamiltoniant}) in the thermodynamic limit.

So far we have used the ground state energy to study the contribution of the inhomogeneous term of the $T-Q$ relation (\ref{inhTQ}). We shall further investigate the contribution of the inhomogeneous term through the first excited state (in Appendix A) and the higher order conserved charges such as the momentum and the second logarithm derivative of the transfer matrix (in Appendix B). The results show that the contribution of the inhomogeneous term can be neglected in the thermodynamic limit.

\section{Thermodynamic limit}
\setcounter{equation}{0}

Now, we consider the thermodynamic limit of the system. For convenience, we define the counting function
\begin{equation}\label{countingt}
  Z_t(x)=\frac{1}{2\pi} \left[ \frac{\eta x}{N} + \theta_1(x) - \frac{1}{N} \sum^M_{k=1} \theta_2(x - x_k)  \right].
\end{equation}
In the thermodynamic limit, $N \rightarrow \infty$, $M \rightarrow \infty$ and $N /M$ takes the finite value. Taking the derivative of Eq. (\ref{countingt}) with respect to $x$, we obtain
\begin{eqnarray}\label{densityt}
\frac{d Z_t(x)}{d x}&=& \frac{\eta}{2\pi N} +a_1(x) - \int^Q_{-Q} a_2(x-y) \rho(y) dy \nonumber  \\
&\equiv &  \rho(x)+\rho^h(x),
\end{eqnarray}
and
\begin{equation}\label{afunction}
 a_m(x)= \frac{1}{2 \pi} \frac{\partial \theta_m(x)}{\partial x}=\frac{\eta}{2 \pi} \frac{\sinh(m \eta)}{\cosh(m \eta)-\cos(\eta x)},
\end{equation}
where $Q$ is chosen as $\pi/\eta$, $\rho(x)$ and $\rho^h(x)$ are the densities of particles and holes, respectively.
For the arbitrary periodic function $f(x)$, $x\in [-Q,Q]$, we introduce the Fourier transformation
\begin{eqnarray}
  \tilde{f}(k) &=& \int^Q_{-Q} f(x) e^{-ik \frac{\pi}{Q} x} dx , \label{Ft1} \\
  f(x) &=& \frac{1}{2Q} \sum^{\infty}_{k=-\infty} \tilde{f}(k) e^{ik \frac{\pi}{Q} x} , ~~~~~~~~~k= \cdots,-2,-1,0,1,2,\cdots. \label{Ft2}
\end{eqnarray}
Taking the Fourier transformation of Eq. (\ref{densityt}), we obtain
\begin{equation}
  \tilde{\rho}(k) +  \tilde{\rho}^h(k)= \frac{1}{N}\delta_{k,0} +\tilde{a}_1(k)-\tilde{a}_2(k) \tilde{\rho}(k),
\end{equation}
where
\begin{equation}\label{afunctionF}
  \tilde{a}_m(k)=e^{-m \eta |k|}.
\end{equation}
Then we have
\begin{equation}\label{dst}
  \tilde{\rho}(k)= \frac{1}{e^{\eta |k|}+e^{- \eta |k|}}  +\frac{1}{N} \frac{ \delta_{k,0} }{ 1+e^{-2\eta |k|} } -\frac{ \tilde{\rho}^h(k) }{ 1+e^{-2\eta |k|} }.
\end{equation}
In the thermodynamic limit, the eigenvalue (\ref{eigenvaluet1}) can be expressed by the density of particles as
\begin{equation}\label{eigenvaluet2}
   E_{hom}=E = - \frac{8 \pi}{\eta} N \sinh \eta \int^Q_{-Q} a_1(x) \rho (x) dx + N \cosh \eta +2\sinh \eta.
\end{equation}

For the even $N$ case, we have $M=\frac{N}{2}$ at the ground state. Thus the following equation must hold
\begin{equation}\label{dst111111dsf}
  \frac{M}{N} = \int^Q_{-Q} \rho(x) dx = \tilde{\rho}(0) =\frac{1}{2}.
\end{equation}
From Eqs. (\ref{dst}) and (\ref{dst111111dsf}), we find that at the ground state, there exists one hole at $x_0 \in[-\frac{\pi}{\eta},\frac{\pi}{\eta}]$.
The density of holes is given by
\begin{equation}\label{hole}
  \rho^h(x)=\frac{1}{N} \delta (x-x_0).
\end{equation}
With the Fourier transformation, we have
\begin{equation}\label{holeF}
  \tilde{\rho}^h(k)=\frac{1}{N} e^{-ik \eta x_0}.
\end{equation}
Thus the solution of (\ref{densityt}) can be derived as
\begin{equation}\label{dst-even}
  \tilde{\rho}(k)= \frac{1}{e^{\eta |k|}+e^{- \eta |k|}}  +\frac{1}{N} \frac{ \delta_{k,0} }{ 1+e^{-2\eta |k|} } -\frac{1}{N} \frac{ e^{-ik \eta x_0} }{ 1+e^{-2\eta |k|} } .
\end{equation}
With the help of Eqs. (\ref{eigenvaluet2}) and (\ref{dst-even}), we obtain
\begin{equation}\label{Et-even}
  E^{even}= e_0 N+e_h(x_0),
\end{equation}
where $e_0$ is exactly the density of ground state energy of the XXZ spin chain with periodic boundary condition
\begin{eqnarray}
e_0 = -8  \sinh \eta \sum^{\infty}_{k=1} \frac{1}{1+e^{2\eta k}} -2 \sinh \eta + \cosh \eta, \label{gEd}
\end{eqnarray}
and $e_h(x_0)$ is the energy carried by one hole as
\begin{eqnarray}
e_h(x_0)= 4 \sinh \eta \sum^{\infty}_{k=-\infty} \frac{e^{ik \eta x_0}}{2 \cosh(\eta k)}. \label{Eh}
\end{eqnarray}
For  the ground state, the position of hole should be put at $x_0=\frac{\pi}{\eta}$ to minimize the energy.
Thus the ground state energy in the thermodynamic limit can be written as
\begin{eqnarray}\label{gEt-even}
  E^{g, even} = e_0 N+e_h(\frac{\pi}{\eta}) .
\end{eqnarray}

For the odd $N$ case, we consider the case that $M=\frac{N+1}{2}$ for the ground state. Thus the following equation must hold
\begin{equation}
  \frac{M}{N} = \int^Q_{-Q} \rho(x) dx = \tilde{\rho}(0) =\frac{1}{2}+\frac{1}{2N}.
\end{equation}
Such a configuration gives that there is no hole and the ground state is completely determined by the density of particles
\begin{equation}\label{dst-odd}
  \tilde{\rho}(k)= \frac{1}{e^{\eta |k|}+e^{- \eta |k|}}  +\frac{1}{N} \frac{ \delta_{k,0} }{ 1+e^{-2\eta |k|} }.
\end{equation}
With the help of Eqs. (\ref{eigenvaluet2}) and (\ref{dst-odd}), we have
\begin{equation}\label{gEt-odd}
  E^{g, odd} = e_0 N.
\end{equation}
with $e_0$ defined as (\ref{gEd}).

From the above calculation, we find that the ground state energy of the antiperiodic XXZ spin chain with even $N$ and that with odd $N$ are different.
This is consistent with the fact that we consider the antiferromagnetic coupling and the massive region of the model (\ref{Hamiltoniant}), i.e., $\Delta=\cosh\eta\geq 1$ with real $\eta$. The values of $e_0$ and $e_h(\pi/\eta)$ have the same order. In the thermodynamic limit, the most contributions come from $e_0N$ and the $e_h(\pi/\eta)$ can be neglected.
Thus the thermodynamic quantities calculated by the density of ground state energy $e_0$ do no depend on the even or odd of $N$. However, in this paper, we focus on the effects induced by the boundary degree of freedom, thus the contribution of $e_h(\pi/\eta)$ can not be neglected. If $\eta \rightarrow 0$, then $e_h(\pi/\eta)\rightarrow 0$.

\section{Thermodynamic limit of the periodic XXZ spin chain}
\setcounter{equation}{0}

In order to study the effects induced by the twisted boundary, now we should study the thermodynamic limit of the XXZ spin chain with periodic boundary condition. The model Hamiltonian reads
\begin{equation}\label{Hamiltonianp}
H_p=\sum^N_{j=1} \left[  \sigma^x_j \sigma^x_{j+1} + \sigma^y_j \sigma^y_{j+1} + \cosh\eta \sigma^z_j \sigma^z_{j+1} \right],
\end{equation}
with the constraint $\sigma_{N+1}^{\alpha} = \sigma_1^{\alpha}$.
We consider the same case that $\eta$ is real, thus the eigenvalues of the Hamiltonian (\ref{Hamiltonianp}) is
\begin{equation} \label{eigenvaluep}
 E_p = -4  \sinh\eta ~\sum^M_{j=1} \frac{\sinh \eta}{\cosh \eta - \cos(\eta x_j)}  + N \cosh \eta,
\end{equation}
where the $M$ Bethe roots $\{x_j\}$ are determined by the BAEs \cite{Takahashi}
\begin{equation}\label{BAEp}
  \frac{\sin^N \frac{\eta}{2} (x_j-i)}{ \sin^N \frac{\eta}{2} (x_j+i) } = -\prod^M_{k=1} \frac{\sin \frac{\eta}{2} (x_j-x_k-2i)}{ \sin \frac{\eta}{2} (x_j-x_k +2i) }.
\end{equation}
Taking the logarithm of Eq. (\ref{BAEp}), we have
\begin{equation}\label{logBAEp}
  N \theta_1(x_j)=2 \pi I_j +\sum^M_{k=1} \theta_2(x_j - x_k),
\end{equation}
where $\{ I_j \}$ are certain integers (half odd integers) for $N-M$ odd ($N-M$ even).
For convenience, we define the counting function
\begin{equation}\label{countp}
  Z_p(x)=\frac{1}{2\pi} \left[   \theta_1(x) - \frac{1}{N} \sum^M_{k=1} \theta_2(x - x_k)  \right] .
\end{equation}
Obviously, $Z_p(x_j)=\frac{I_j}{N} $ corresponds to the Eq. (\ref{logBAEp}) and it will turn to be a continuous function in the thermodynamic limit.
When $N \rightarrow \infty$ and $M \rightarrow \infty$, the distribution of Bethe roots are continuous, i.e., $Z_p(x_j)=Z_p(x)$. Taking the derivative of Eq. (\ref{countp}) with respect to $x$, we obtain
\begin{eqnarray}\label{densityp}
\frac{d Z_p(x)}{d x} &=&a_1(x) - \int^Q_{-Q} a_2(x-y) \rho(y) dy \nonumber \\ &=&  \rho(x)+\rho^h(x),
\end{eqnarray}
where $\rho(x)$ and $\rho^h(x)$ are the densities of the particles and holes, respectively. Taking the Fourier transformation of Eq. (\ref{densityp}),
we obtain
\begin{eqnarray}
  \tilde{\rho}(k) +  \tilde{\rho}^h(k)=\tilde{a}_1(k)-\tilde{a}_2(k) \tilde{\rho}(k).
\end{eqnarray}
Thus the density of particles can be expressed as
\begin{equation}\label{densityt-solutiion-p}
  \tilde{\rho}(k)= \frac{1}{e^{\eta |k|}+e^{- \eta |k|}} -\frac{ \tilde{\rho}^h(k) }{ 1+e^{-2\eta |k|} }.
\end{equation}
In the thermodynamic limit, the energy (\ref{eigenvaluep}) of the periodic XXZ spin chain is
\begin{equation}\label{eigenvaluep2}
   E_p = - \frac{8 \pi}{\eta} N \sinh \eta \int^Q_{-Q} a_1(x) \rho (x) dx + N \cosh \eta .
\end{equation}

For the even $N$, all the Bethe roots are real at the ground state and fill the region $(-\frac{\pi}{\eta}, \frac{\pi}{\eta}]$. Meanwhile, the number of Bethe roots $M=\frac{N}{2}$.
Thus the following equation must hold
\begin{equation}\label{wqeo233}
  \frac{M}{N} = \int^Q_{-Q} \rho(x) dx = \tilde{\rho}(0) =\frac{1}{2},
\end{equation}
which means the magnetization of the ground state is 0 \cite{Takahashi}.
From Eqs. (\ref{densityt-solutiion-p}) and (\ref{wqeo233}), we find that such a configuration is described by $\rho^h(x)=0$ and the density of particles is
\begin{equation}\label{dsp-even}
  \tilde{\rho}(k)= \frac{1}{e^{\eta |k|}+e^{- \eta |k|}}=\frac{1}{2 \cosh (\eta k)}.
\end{equation}
With the help of Eqs. (\ref{eigenvaluep2}) and (\ref{dsp-even}), we have
\begin{equation}\label{gEp-even}
  E^{g,even}_p = e_0 N,
\end{equation}
where $e_0$ is the density of ground state energy of the system defined by Eq. (\ref{gEd}).

For the odd $N$, the ground state of the system (\ref{Hamiltonianp}) is described by $\frac{N-1}{2}$ real Bethe roots in the region $(-\frac{\pi}{\eta}, \frac{\pi}{\eta}]$ and one hole at $x_0 \in[-\frac{\pi}{\eta},\frac{\pi}{\eta}]$. Thus the following equation must hold
\begin{equation}
  \frac{M}{N} = \int^Q_{-Q} \rho(x) dx = \tilde{\rho}(0) =\frac{1}{2}-\frac{1}{2N}.
\end{equation}
In this case, the density of holes is given by Eq.(\ref{hole}).

Then from Eq. (\ref{densityt-solutiion-p}), we obtain the density of particles as
\begin{equation}\label{dsp-odd}
  \tilde{\rho}(k)= \frac{1}{2 \cosh (\eta k)} -\frac{1}{N} \frac{ e^{-ik \eta x_0} }{ 1+e^{-2\eta |k|} }.
\end{equation}
With the help of Eqs.(\ref{eigenvaluep2}) and (\ref{dsp-odd}), we have
\begin{equation}\label{Ep-odd}
  E^{odd}_p = e_0 N+e_h(x_0),
\end{equation}
where $e_h(x_0)$ is the energy carried by one hole defined by Eq. (\ref{Eh}).
At the ground state, $x_0=\frac{\pi}{\eta}$ to minimize the energy.
Thus the ground state energy in the thermodynamic limit can be expressed by
\begin{equation}\label{gEp-odd}
  E^{g,odd}_p = e_0 N+e_h(\frac{\pi}{\eta}).
\end{equation}

Again, we find that the ground state energy of the periodic XXZ spin chain with even $N$ and that with odd $N$ are different. In the thermodynamic limit, comparing with $e_0 N$, the $e_h(\pi/\eta)$ is a small quantity and can be neglected. The thermodynamic behavior of the system with even $N$ and those with odd $N$ obtained by the density of ground state energy $e_0$ are the same.

Comparing the relations (\ref{gEp-even}) and (\ref{gEt-even}), (\ref{gEp-odd}) and (\ref{gEt-odd}), we find that the parity of $N$ of the antiperiodic XXZ spin chain and the parity of $N$ of the periodic XXZ spin chain are reversed. That is to say, the ground state energy of the periodic XXZ spin chain with even $N$ equals to that of the antiperiodic XXZ spin chain with odd $N$, while the ground state energy of the periodic XXZ spin chain with odd $N$ equals to that of the antiperiodic XXZ spin chain with even $N$. This is because of the existence of the twisted bond.

\section{Twisted boundary energy}
\setcounter{equation}{0}

\begin{figure}[!htp]
    \begin{minipage}[t]{0.5\linewidth}
    \centering
    \includegraphics[height=4cm,width=8cm]{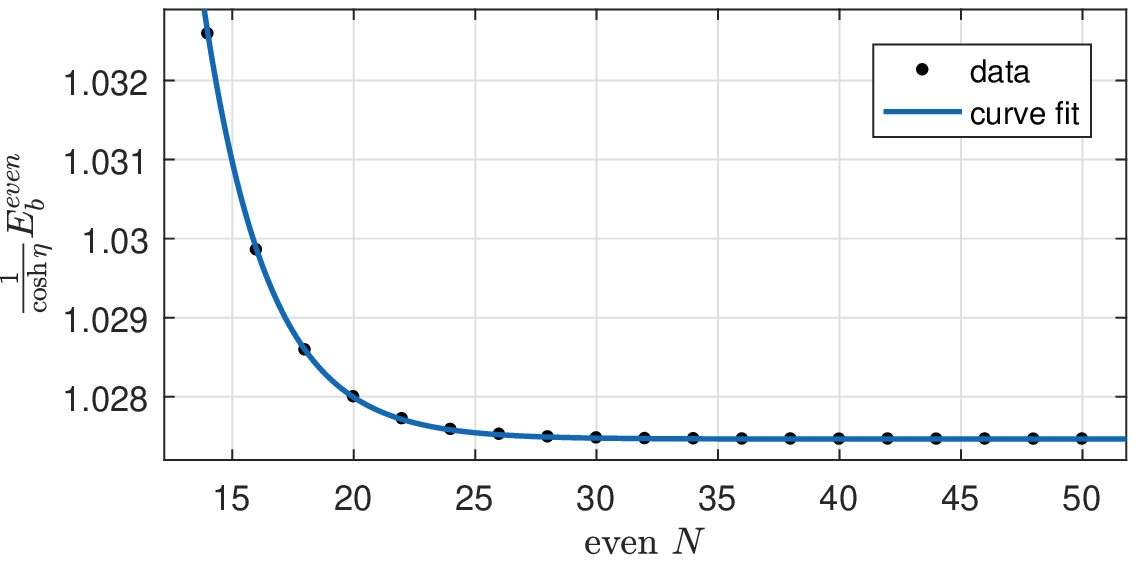}
    \caption*{(a)}
    \end{minipage}
    \begin{minipage}[t]{0.6\linewidth}
    \centering
    \includegraphics[height=4cm,width=8cm]{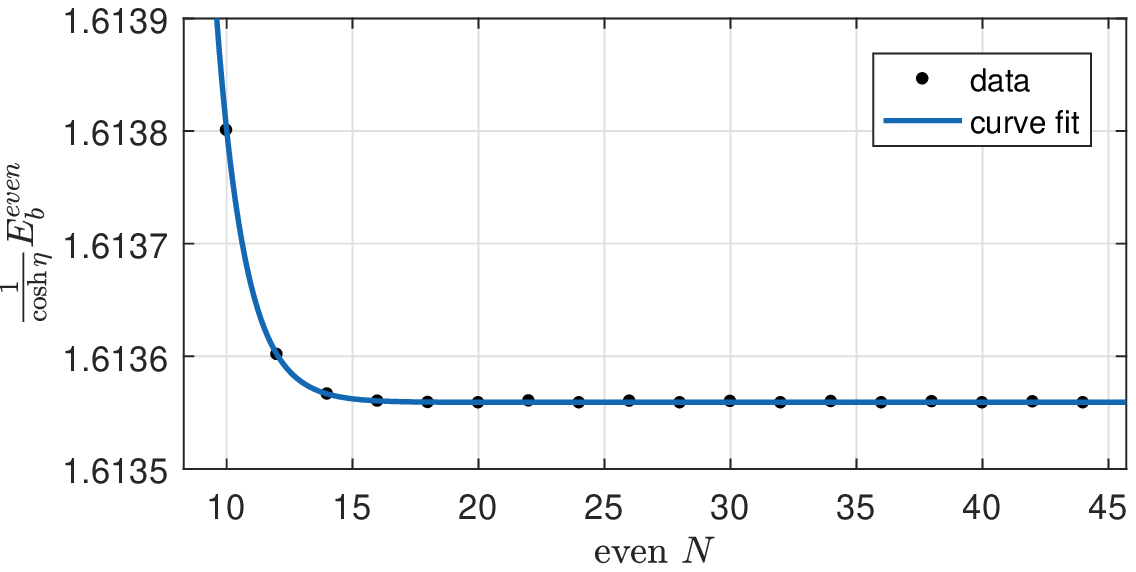}
    \caption*{(b)}
    \end{minipage}
    \begin{minipage}[t]{0.5\linewidth}
    \centering
    \includegraphics[height=4cm,width=8cm]{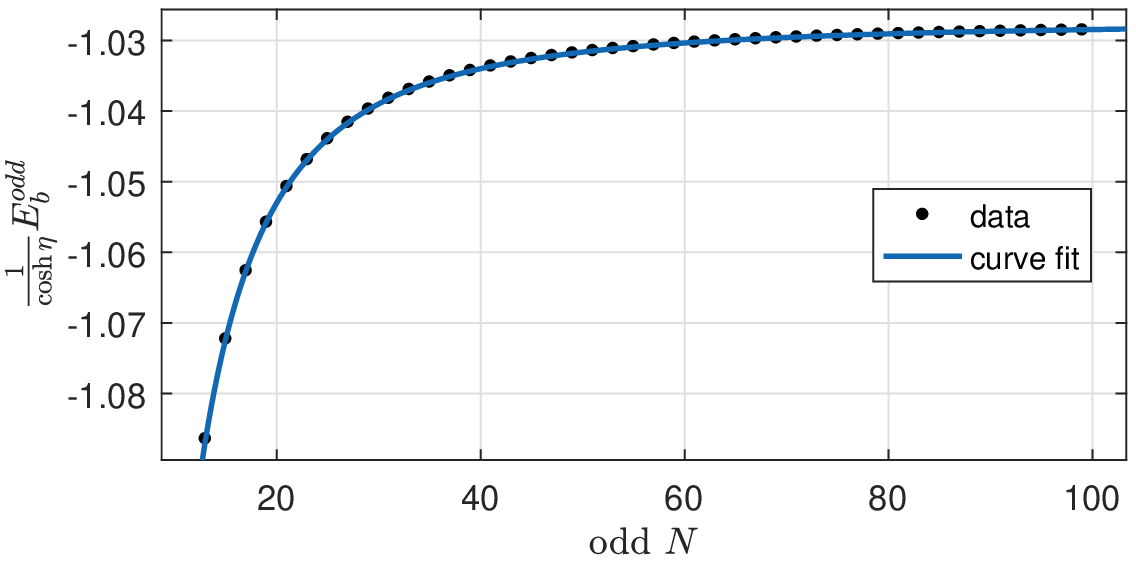}
    \caption*{(c)}
    \end{minipage}
    \begin{minipage}[t]{0.6\linewidth}
    \centering
    \includegraphics[height=4cm,width=8cm]{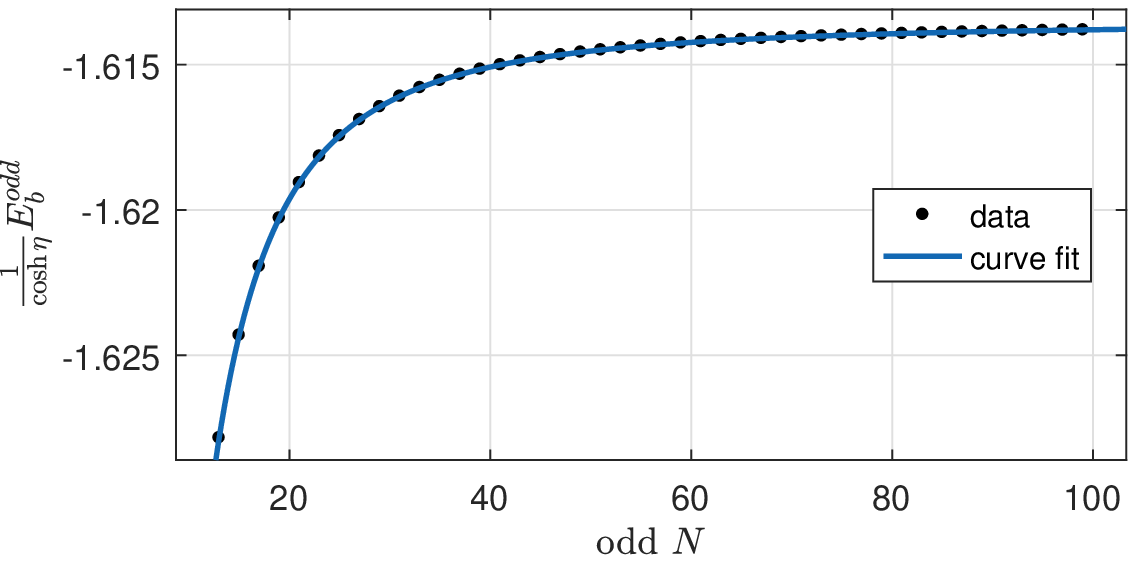}
    \caption*{(d)}
    \end{minipage}
    \caption{The twisted boundary energies $E_b$ versus the system size $N$.
The data in (a) and (b) can be fitted as $\frac{1}{\cosh \eta}E^{even}_b=a_3 e^{b_3N}+c_3$, where (a) $\eta=2$, $a_3=1.028$, $b_3=-0.3787$ and $c_3=1.027$;
(b) $\eta=3$, $a_3=1.461$, $b_3=-0.8706$ and $c_3=1.614$. Due to the fact $b_3<0$, when the system size $N$ tends to infinity, $c_3$ is the corresponding twisted boundary energy.
The data in (c) and (d) can be fitted as $\frac{1}{\cosh \eta}E^{odd}_b =a_4 N^{b_4}+c_4$, where (c) $\eta=2$, $a_4=-8.696$, $b_4=-1.945$ and $c_4=-1.027$; (d) $\eta=3$, $a_4=-2.342$, $b_4=-1.988$ and $c_4=-1.614$.
Due to the $b_4<0$, when the system size $N$ tends to infinity, $|c_4|$ is the corresponding twisted boundary energy.
}\label{fig-Eb}
\end{figure}
The twisted boundary energy is a physical quantity to measure the effects induced by twisted boundary at the ground state, which is defined as
\begin{equation}\label{Eb}
  E_b=|E^g-E^g_p|,
\end{equation}
which is a function of the crossing parameter $\eta$. The symbol of absolute value in Eq. (\ref{Eb}) is used because that $E^{g} > E^g_p$ for even $N$, while $E^g < E^g_p$ for odd $N$.
From Eqs. (\ref{gEp-even}) and (\ref{gEt-even}), we find that $E^{g, even}-E^{g, even}_p$ equals to the twisted boundary energy for even $N$
\begin{eqnarray}\label{Eb-e111o}
  E^{even}_b(\eta)=E^{g, even}(\eta)-E^{g, even}_p(\eta)=E_b(\eta) = 4\sinh \eta \sum^{\infty}_{k=1} \frac{\cos(k \pi)}{\cosh(\eta k)} +2\sinh \eta.
\end{eqnarray}
From Eqs. (\ref{gEp-odd}) and (\ref{gEt-odd}), we find that $E^{g, odd}-E^{g, odd}_p$ equals to the minus of twisted boundary energy for odd $N$
 \begin{eqnarray}\label{Eb-qqeo}
  E^{odd}_b(\eta) =E^{g, odd}(\eta)-E^{g, odd}_p(\eta) =- E_b(\eta).
\end{eqnarray}
Therefore, the twisted boundary energy $E^{even}_b$ with even $N$ equals to the minus of twisted boundary energy $E^{odd}_b$ with odd $N$
 \begin{eqnarray}\label{Eb-eo}
  E^{odd}_b(\eta) = - E^{even}_b(\eta).
\end{eqnarray}
The twisted boundary energies with $\eta=2$ and $\eta=3$ are derived as
\begin{eqnarray}
&& \frac{1}{\cosh 2} E^{even}_b (2) = -\frac{1}{\cosh 2} E^{odd}_b (2)=1.02746,\nonumber\\
&& \frac{1}{\cosh 3} E^{even}_b (3) =-\frac{1}{\cosh 3} E^{odd}_b (3) = 1.61356 \label{Eb-eta3}.
\end{eqnarray}

Now, we check the above results by the DMRG method.
The twisted boundary energies for various system-size $N$ obtained by DMRG are shown in Fig. \ref{fig-Eb}.
For the even $N$ case, the data in Fig. \ref{fig-Eb}(a) and Fig. \ref{fig-Eb}(b) can be fitted as
\begin{equation}
\frac{1}{\cosh \eta}E^{even}_b=a_3 e^{b_3N}+c_3.
\end{equation}
Due to the fact $b_3<0$, when the system size $N$ tends to infinity, $c_3$ should be the twisted boundary energy, $c_3=E_b$. The DMRG results are
\begin{eqnarray}
&& c_3 = 1.027, \qquad {\rm for} \qquad \eta=2, \nonumber \\
&& c_3 = 1.614, \qquad {\rm for} \qquad \eta=3,
\end{eqnarray}
which are highly consistent with the analytical results (\ref{Eb-eta3}).

For the odd $N$ case, the data in Fig. \ref{fig-Eb}(c) and Fig. \ref{fig-Eb}(d) can be fitted as
\begin{equation}
\frac{1}{\cosh \eta} E^{odd}_b=a_4 N^{b_4}+c_4.
\end{equation}
Due to the fact $b_4<0$, when the system size $N$ tends to infinity, $|c_4|$ should be the twisted boundary energy, $|c_4|=E_b$. The DMRG results are
\begin{eqnarray}
&& c_4 = -1.027, \qquad {\rm for} \qquad \eta=2, \nonumber \\
&& c_4 = -1.614, \qquad {\rm for} \qquad \eta=3,
\end{eqnarray}
which are also highly consistent with the analytical results (\ref{Eb-eta3}).

Now, we consider the degenerate case. When $\eta=0$, the antiperiodic XXZ spin chain degenerates into the isotropic XXX spin chain
with the antiperiodic boundary conditions. From Eq. (\ref{Eh}), we have
\begin{eqnarray}\label{evenEtN112}
\lim_{\eta\rightarrow 0}  e_h(\frac{\pi}{\eta})=0.
\end{eqnarray}
Thus the parity of $N$ vanishes and the ground state energy reads
\begin{equation}
 E^{g}_{XXX}= \lim_{\eta\rightarrow 0}  \frac{e_0(\eta)N}{\cosh \eta}  =( 1-4 \ln2 ) N.
\end{equation}
The ground state energy of the periodic XXX spin chain is
\begin{equation}
E^{g}_{pXXX}= ( 1-4 \ln2) N.
\end{equation}
Therefore, the twisted boundary energy of the antiperiodic XXX spin chain is zero.

However, we can further to calculate the first excitation energy of this model base on the reduced $T-Q$ relation (\ref{homTQ}), and the details are shown in the Appendix A. The results  indicate that the low lying excitation can also be obtained by the reduced $T-Q$ relation (\ref{homTQ}) when $N \rightarrow\infty$.

\section{Conclusions}

In this paper, we have studied the thermodynamic limits of the spin-$\frac{1}{2}$ XXZ  chain both with the antiperiodic and the periodic boundary conditions. We find that
due to the twisted bond, the ground state energy of the antiperiodic XXZ spin chain with even $N$ equals to that of the periodic XXZ spin chain with odd $N$, while the ground state energy of the antiperiodic XXZ spin chain with odd $N$ equals to that of the periodic XXZ spin chain with even $N$. We also find that the contribution of the inhomogeneous term in the $T-Q$ relation of the antiperiodic XXZ spin chain for the ground state can be neglected when the system-size $N$ tends to the infinity. By using the reduced BAEs, we study the twisted boundary energy and show that
the twisted boundary energy $E^{even}_b$ of the system with even $N$ differs from the one $E^{odd}_b$ with odd $N$ by a minus sign. We check these results by the DMRG method,  which leads to that  the analytical results and the numerical ones agree with each other very well. We also study the first excitation energy based on the reduced $T-Q$ relation in Appendix A.

\section*{Acknowledgments}

The financial supports from the National Program
for Basic Research of MOST (Grant Nos. 2016YFA0300600 and
2016YFA0302104), the National Natural Science Foundation of China
(Grant Nos. 11434013, 11425522, 11547045, 11774397, 11775178 and 11775177), the Major Basic Research Program of Natural Science of Shaanxi Province
(Grant Nos. 2017KCT-12, 2017ZDJC-32) and the Strategic Priority Research Program of the Chinese
Academy of Sciences, and the Double First-Class University Construction Project of Northwest University
are gratefully acknowledged.  Z. Xin and Y. Qiao are also partially supported by the NWU graduate student innovation funds (Grant Nos. YZZ15088, YYB17003).
We would like to thank F. Wen, P. Sun and X. Xu for their useful discussions.

\section*{Appendix A: The first excitation energy}
\setcounter{equation}{0}
\renewcommand{\theequation}{A.\arabic{equation}}

\begin{figure}[!htp]
    \begin{minipage}[t]{0.5\linewidth}
    \centering
    \includegraphics[height=4.5cm,width=8cm]{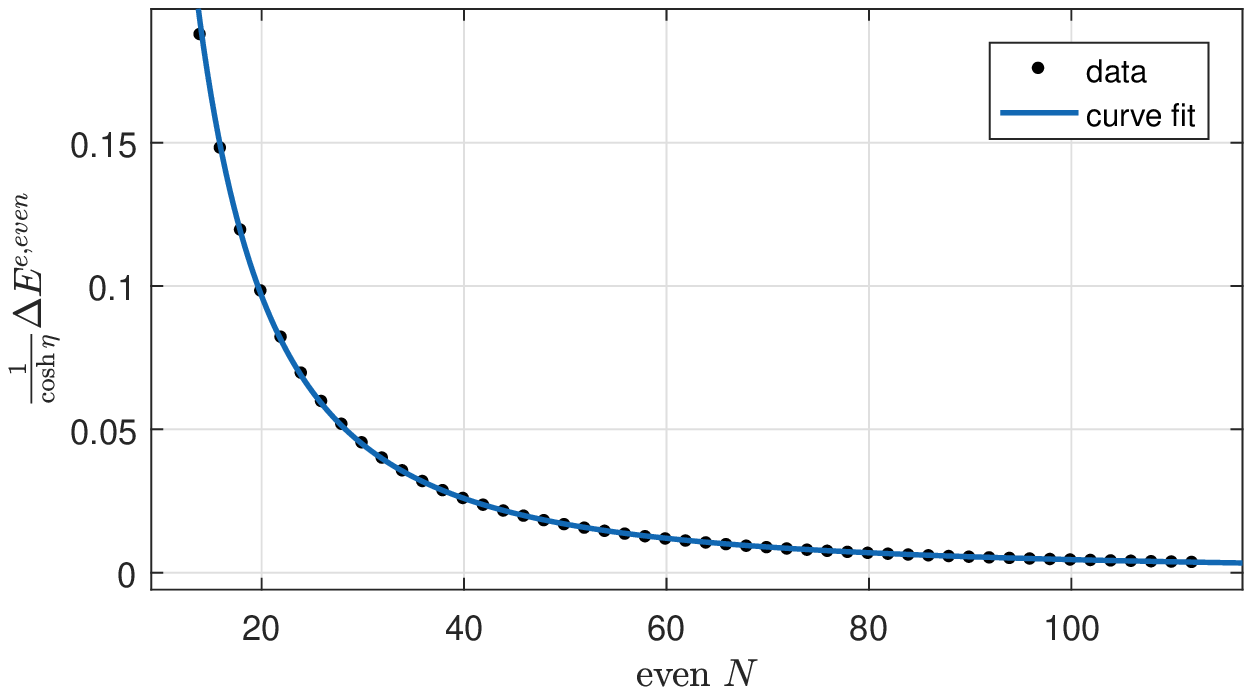}
    \caption*{(a)}
    \end{minipage}
    \begin{minipage}[t]{0.5\linewidth}
    \centering
    \includegraphics[height=4.5cm,width=8cm]{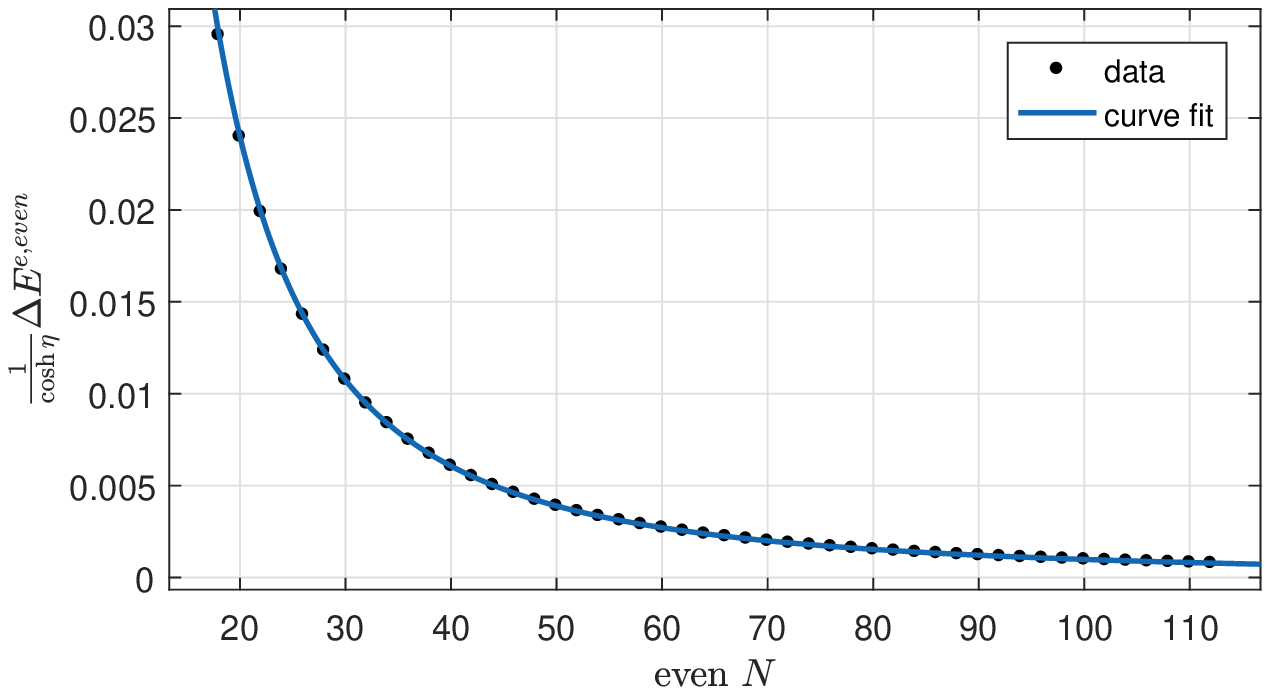}
    \caption*{(b)}
    \end{minipage}
    \caption{
    The first excitation energy $\Delta E^{e,even}$ versus the even system-size $N$.
    The data can be fitted as $\frac{1}{\cosh \eta} \Delta E^{e,even}=a_5 N^{b_5}$.
    Here (a) $\eta=2$, $a_5=28.45$ and $b_5=-1.898$; (b)$\eta=3$, $a_5=9.138$ and $b_5=-1.984$;
    Due to the fact $b_5<0$, the first excitation energy $\Delta E^{e,even}$ tends to zero when $N \rightarrow \infty$.
}\label{eta3evenN}
\end{figure}

In this appendix, we calculate the first excitation energy of the XXZ spin chain with the antiperiodic boundary condition.
In the even $N$ case, one can obtain the low lying excited energy $\Delta E^{even}$ from Eq.(\ref{Et-even}) and Eq.(\ref{gEt-even}) as
\begin{equation}%\label{}
  \Delta E^{even}(x_0)=E^{even}-E^{g,even}=e_h(x_0)-e_h(\frac{\pi}{\eta}).
\end{equation}
Obviously, $\Delta E^{even}$ will turn to be a continuous function in the thermodynamic limit. At the first excited state, the position $x_0$ of hole should tend to $\frac{\pi}{\eta}$  to minimize the $\Delta E^{even}(x_0)$. So the first excitation energy, denoted by $\Delta E^{e,even}$, can be obtained as
\begin{equation}%\label{}
\Delta E^{e,even}=\lim_{x_0\rightarrow \frac{\pi}{\eta}} \Delta E^{even}(x_0) \rightarrow 0.
\end{equation}
On the other hand, the first excitation energy for various even system-size $N$ obtained by DMRG are shown in Fig.\ref{eta3evenN}.
The data  can be fitted as
\begin{equation}
  \frac{1}{\cosh \eta} \Delta E^{e,even}=a_5 N^{b_5}.
\end{equation}
Due to the fact $b_5<0$, the value of $\Delta E^{e,even}$ tends to zero when the system size $N$ tends to infinity, which is highly consistent with the analytical result.
This implies that the low lying excitation energy will be continuous in the even $N$ case.

\begin{figure}[!htp]
    \begin{minipage}[t]{0.5\linewidth}
    \centering
    \includegraphics[height=4.5cm,width=8cm]{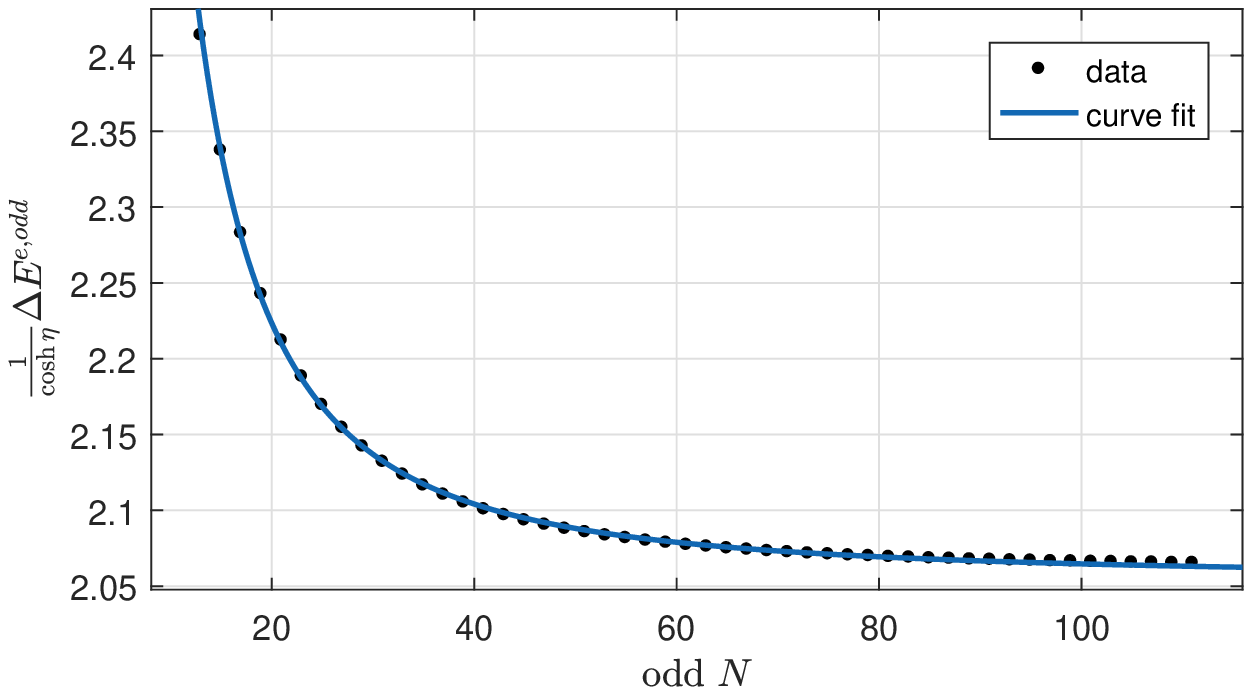}
    \caption*{(a)}
    \end{minipage}
    \begin{minipage}[t]{0.5\linewidth}
    \centering
    \includegraphics[height=4.5cm,width=8cm]{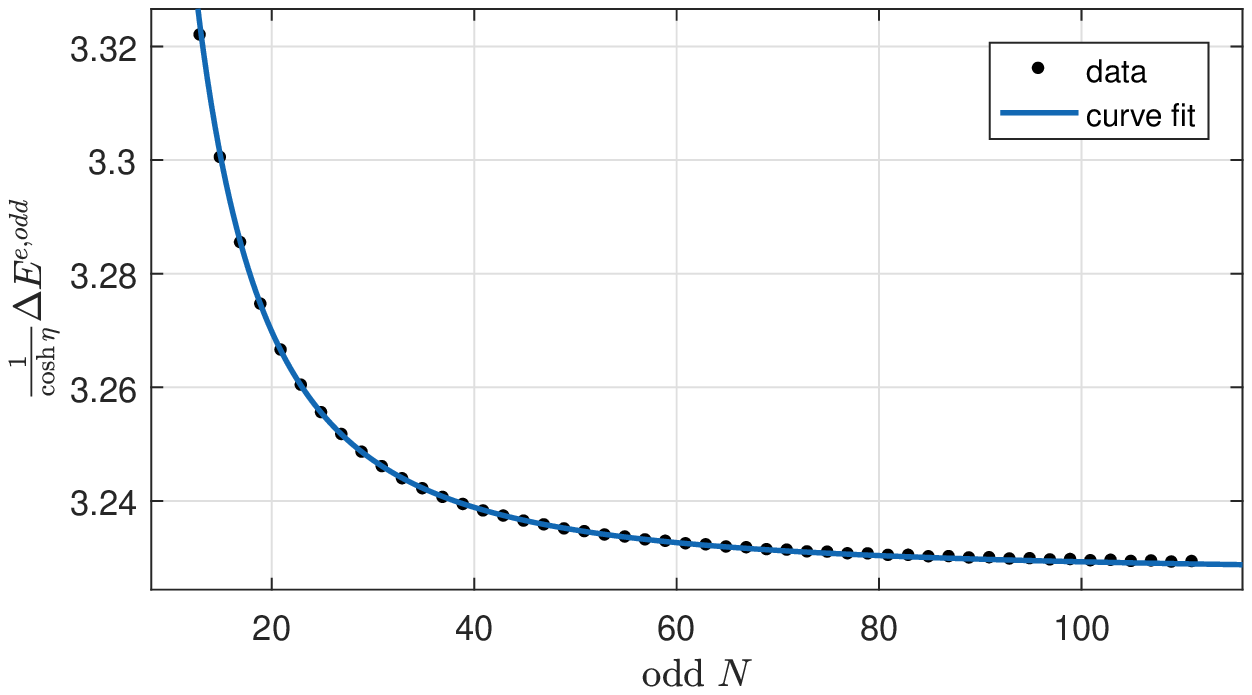}
    \caption*{(b)}
    \end{minipage}
    \caption{
    The first excitation energy $ \Delta E^{e,odd}$ versus the odd system-size $N$.
    The data can be fitted as $\frac{1}{\cosh \eta} \Delta E^{e,odd}=a_6 N^{b_6} +c_6$.
    Here (a) $\eta=2$, $a_6=34.64$, $b_6=-1.778$ and $c_6=2.055$; (b) $\eta=3$, $a_6=11.41$, $b_6=-1.866$ and $c_6=3.227$.
    Due to the fact $b_6<0$,  when $N$ tends to infinity, $c_6$ is the corresponding  first excitation energy $\frac{1}{\cosh \eta} \Delta E^{e,odd}$.}
\label{eta3oddN}
\end{figure}

In the odd $N$ case, the simplest excitation is the hole excitation, with $M'=\frac{N-1}{2}$. Moreover the equation  holds
\begin{equation}
  \frac{M'}{N}=\int^Q_{-Q} \rho(x) dx =\frac{1}{2} -\frac{1}{2N}.
\end{equation}
Such a configuration is described with two holes puting at $x_1$ and $x_2$ $\in[-\frac{\pi}{\eta},\frac{\pi}{\eta}]$,
the density of holes is
\begin{equation}
  \rho^h(x)=\frac{1}{N} \left( \delta (x-x_1) + \delta (x-x_2) \right).
\end{equation}
Using the similar method in the section 4, we can obtain the higher energy $ E^{odd}$ as
\begin{equation}
  E^{odd}=e_0 N+e_h(x_1)+e_h(x_2).
\end{equation}
From the above calculation and relation (\ref{gEt-odd}), we obtain the excitation energy $\Delta E^{odd}$ as
\begin{equation}
  \Delta E^{odd}(x_1,x_2)=E^{odd}-E^{g,odd}=e_h(x_1)+e_h(x_2).
\end{equation}
In the first excited state, $x_1=x_2=\frac{\pi}{\eta}$ to minimize the $\Delta E^{odd}$. So we obtain the first excitation energy as
\begin{equation}
 \Delta E^{e,odd}(\eta)=2e_h(\frac{\pi}{\eta}).
\end{equation}
Here the first excitation energy with $\eta=2$ and $\eta=3$ are derived as
\begin{eqnarray}\label{ana}
  \frac{1}{\cosh 2}\Delta E^{e,odd}(2) &=& 2.05492,  \nonumber \\
  \frac{1}{\cosh 3}\Delta E^{e,odd}(3) &=& 3.22712.
\end{eqnarray}
On the other hand, the first excitation energy for various odd system-size $N$ obtained by DMRG are shown in Fig.\ref{eta3oddN}.
The resulting data  can be fitted as
\begin{equation}
  \frac{1}{\cosh \eta} \Delta E^{e,odd}=a_6 N^{b_6} +c_6.
\end{equation}
Due to the fact $b_6<0$, when the system size $N$ tends to infinity, $c_6 $ should be the first excitation energy, $c_6= \frac{1}{\cosh \eta} \Delta E^{e,odd}$.
The DMRG results are
\begin{eqnarray}
  c_6 &=& 2.055  \qquad {\rm for} \qquad \eta=2, \nonumber \\
  c_6 &=& 3.227  \qquad {\rm for} \qquad \eta=3,
\end{eqnarray}
which are highly consistent with the analytical results (\ref{ana}). It also implies that the elementary excitations possess a finite gap in the odd $N$ case.

%the momentum and the second logarithm derivative
\section*{Appendix B: Finite-size effects of other conserved charges}
\setcounter{equation}{0}
\renewcommand{\theequation}{B.\arabic{equation}}

In this appendix, we calculate the finite-size corrections for the  momentum and the second order logarithm derivative of the transfer matrix.
As for the momentum,  the corresponding operator can be expressed in terms of the transfer matrix  as
\begin{equation}\label{0dt}
  H^{(0)}=\ln t(0)=\ln \left[ \sigma^x_1 P_{1,N} \cdots  P_{1,2} \right].
\end{equation}
The eigenvalue $E^{(0)}$ of $H^{(0)}$ (\ref{0dt}) is then given in terms of the associated Bethe roots as
\begin{equation}\label{E0}
  E^{(0)}= \ln \Lambda(0)=\sum^N_{j=1} \left[ \ln \left(\sin \frac{\eta}{2}(x_j-i) \right) - \ln \left(\sin \frac{\eta}{2}(x_j+i) \right)  \right],
\end{equation}
where the transformation $\lambda_j= \frac{\eta}{2} ( i x_j -1)$ is used in (\ref{inhTQ}).
We know that $t^{2N}(0)=1$, so $\Lambda^{2N}(0)=1$. Then, the following relations can be obtained
\begin{eqnarray}
  \Lambda(0) &=& e^{\frac{k}{N}\pi i}, \\
  E^{(0)} &=&  \frac{k}{N}\pi i,  \quad \quad k=-N \cdots N.
\end{eqnarray}

Following the method in section 3, we define
\begin{equation}\label{redE0}
  E^{(0)}_{hom}= \ln \Lambda_{hom}(0)=\sum^M_{j=1} \left[ \ln \left(\sin \frac{\eta}{2}(x_j-i) \right) - \ln \left(\sin \frac{\eta}{2}(x_j+i) \right)  \right].
\end{equation}
Then the contribution of the inhomogeneous term to the $E^{(0)}$ can be defined as
\begin{equation}\label{cE0}
  E^{(0)}_{inh}\equiv E^{(0)}_{hom}-E^{(0)} ,
\end{equation}
where the $E^{(0)}$ can be calculated by BAEs (\ref{Inhomogeneous BAE}) and relation (\ref{E0}) or by direct diagonalization of the $H^{(0)}$ (\ref{0dt}).
$E^{(0)}_{hom}$ can be obtained by Eqs.(\ref{logBAEt}) (or the reduced BAEs (\ref{BAEt1})) and (\ref{redE0}).

\begin{figure}[!htp]
    \begin{minipage}[t]{0.5\linewidth}
    \centering
    \includegraphics[height=4.5cm,width=8cm]{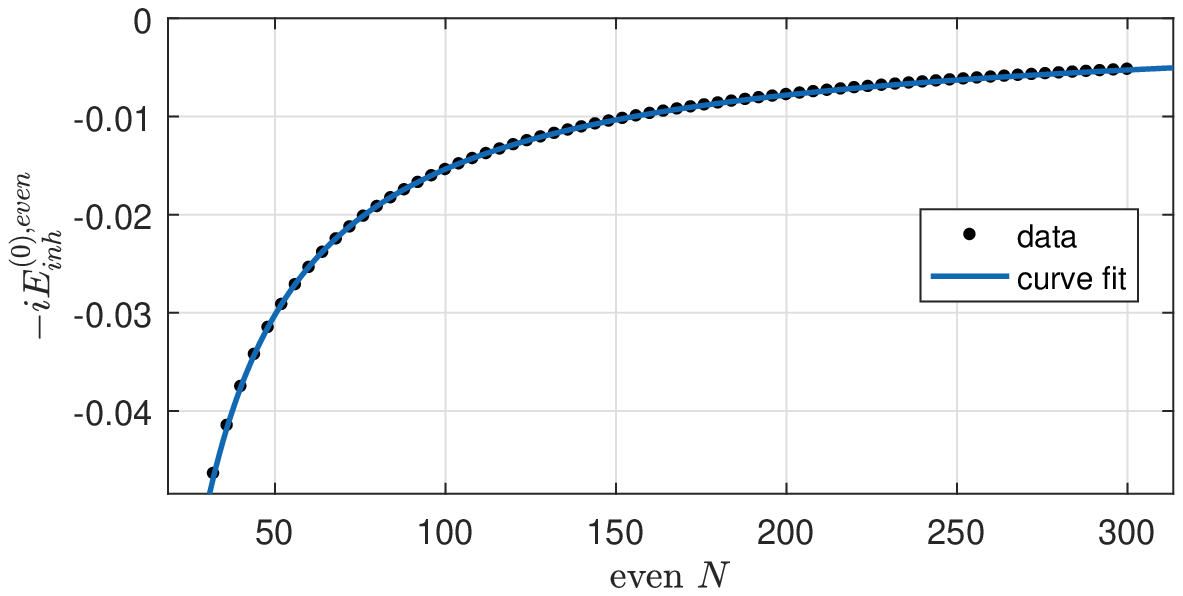}
    \caption*{(a)}
    \end{minipage}
    \begin{minipage}[t]{0.5\linewidth}
    \centering
    \includegraphics[height=4.5cm,width=8cm]{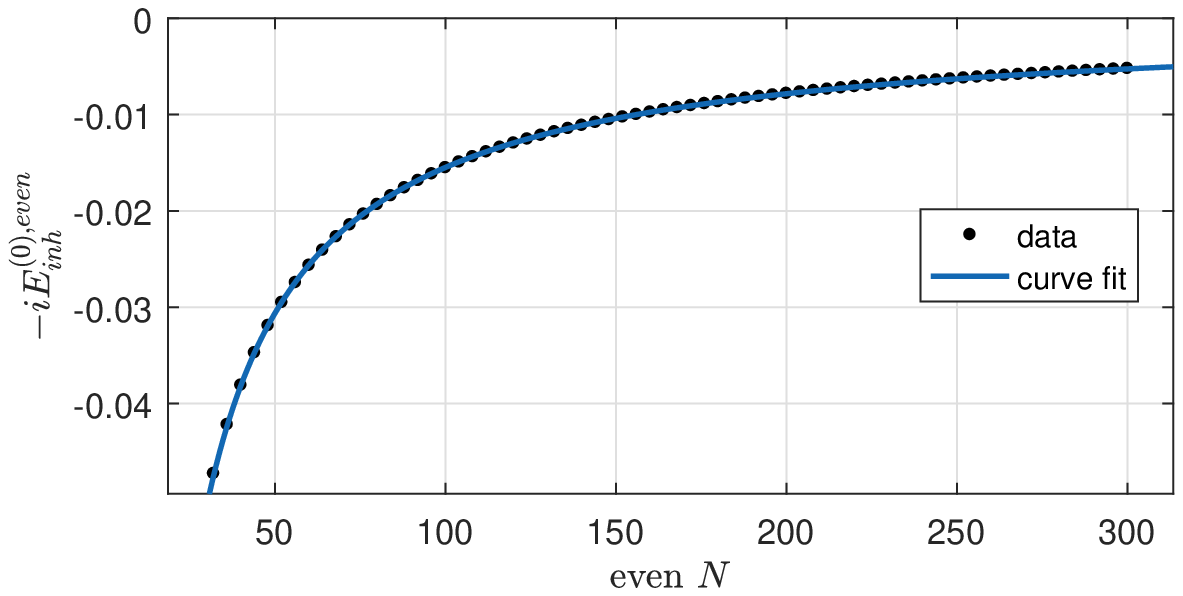}
    \caption*{(b)}
    \end{minipage}
    \caption{The contribution of the inhomogeneous term to the momentum $ E^{(0),even}_{inh}$ versus the even system-size $N$.
The data can be fitted as $ E^{(0),even}_{inh}(N)=i a_7 N^{b_7}$.
Here (a) $\eta=2$, $a_7=-1.367$ and $b_7=-0.9746$; (b) $\eta=3$, $a_7=-1.432$ and $b_7=-0.983$. Due to the fact $b_7< 0$ ($b_7\simeq -1$), the contribution of the inhomogeneous term tends to zero when the $N \rightarrow \infty$.}\label{fig-FC0j-even}
\end{figure}

Now, we calculate the $E^{(0)}_{inh}$ corresponding to the ground state of Hamiltonian (\ref{Hamiltoniant}). In the even $N$ case, one can find that there are two eigenvalues $E^{(0),even}=\pm \frac{\pi}{2}i$ corresponding to the two double degenerate ground states of Hamiltonian (\ref{Hamiltoniant}).
Without loss of generality, we consider the one ground state $\mid g^{even}>$ of the two double degenerate ground states which satisfies the relation $E^{(0),even}=<g^{even} \mid H^{(0)} \mid g^{even}>=\frac{\pi}{2}i$  for any even $N$.
Here the $E^{(0),even}_{hom}$ can be obtained by Eqs.(\ref{redE0}) and   (\ref{logBAEt}) which the Bethe roots determined by the quantum number (\ref{qnumber-t-g-even}).
From Eq. (\ref{cE0}), the results can be shown in Fig.\ref{fig-FC0j-even}.
From the fitting, we find that $E^{(0),even}_{inh}$ and even $N$ satisfy the power law
\begin{equation}
 E^{(0),even}_{inh}(N)=i a_7 N^{b_7}.
\end{equation}
Due to the fact that $b_7<0$ ($b_7\simeq -1$),
we find that there is a correction of the order 1/N in the momentum and the contribution of the inhomogeneous term can be neglected when $N\rightarrow\infty$.

In the odd $N$ case, one can find that there are two eigenvalues $E^{(0),odd}=0, \pi i$ corresponding to the two double degenerate ground states of Hamiltonian (\ref{Hamiltoniant}).
Without loss of generality, we consider the one ground state $\mid g^{odd}>$ of the two double degenerate ground states which satisfies the relation $E^{(0),odd}=<g^{odd} \mid H^{(0)} \mid g^{odd}>=0$  for any odd $N$.
Here the $E^{(0),odd}_{hom}$ can be obtained by Eqs.(\ref{redE0}) and  (\ref{logBAEt}) which the Bethe roots determined by the quantum number (\ref{qnumber-t-g-odd}). It is easy to check  $E^{(0),odd}_{hom}=0$. Therefore we can conclude that
\begin{equation}
 E^{(0),odd}_{inh}(N)=0,
\end{equation}
namely,  there is no correction for the total momentum in the odd $N$ case.

The higher conserved charge can be obtained by the second order logarithm derivative of the transfer matrix $t(u)$ (\ref{traM})
\begin{eqnarray}\label{2dt}
  H^{(2)} &=& 2i \sinh^2 \eta \frac{\partial^2 \ln t(u)}{\partial u^2} \mid_{u=0,\{\theta_j=0\}} +2N i  \nonumber \\
   &=& \sum^N_{j=1}  -\cosh \eta \sigma^x_{j}\sigma^y_{j+1} \sigma^z_{j+2}  + \cosh \eta \sigma^y_{j} \sigma^x_{j+1} \sigma^z_{j+2}  - \sigma^y_{j}\sigma^z_{j+1} \sigma^x_{j+2}   \nonumber \\
   &&  + \cosh \eta \sigma^z_{j} \sigma^y_{j+1} \sigma^x_{j+2} -\cosh \eta \sigma^z_{j}\sigma^x_{j+1} \sigma^y_{j+2}  +  \sigma^x_{j} \sigma^z_{j+1} \sigma^y_{j+2} ,
\end{eqnarray}
and the antiperiodic boundary condition reads
\begin{equation}
  \sigma_{N+k}^{\alpha} = \sigma_{k}^{x} \sigma_{k}^{\alpha} \sigma_{k}^{x}, \quad  k=1,~2.
\end{equation}
The eigenvalue $E^{(2)}$ of the $H^{(2)}$ (\ref{2dt}) is then expressed in terms of the associated Bethe roots as
\begin{eqnarray}\label{E2}
  E^{(2)} &=& 2i \sinh^2 \eta \frac{\partial^2 \ln \Lambda (u)}{\partial u^2} \mid_{u=0,\{\theta_j=0\}} +2N i  \nonumber \\
   &=& 2i \sinh^2 \eta \sum^N_{j=1}\left[ \cot^2\frac{\eta}{2}(x_j-i) - \cot^2\frac{\eta}{2}(x_j+i)  \right] .
\end{eqnarray}

Following the method in section 3, we define
\begin{eqnarray}\label{redE2}
  E^{(2)}_{hom} &=& 2i \sinh^2 \eta \frac{\partial^2 \ln \Lambda_{hom} (u)}{\partial u^2} \mid_{u=0,\{\theta_j=0\}} +2N i  \nonumber \\
   &=& 2i \sinh^2 \eta \sum^M_{j=1}\left[ \cot^2\frac{\eta}{2}(x_j-i) - \cot^2\frac{\eta}{2}(x_j+i)  \right] .
\end{eqnarray}
The contribution of the inhomogeneous term to the $E^{(2)}$ can be defined as
\begin{equation}\label{cE2}
  E^{(2)}_{inh}\equiv E^{(2)}_{hom}-E^{(2)} ,
\end{equation}
where the $E^{(2)}$ can be calculated by BAEs (\ref{Inhomogeneous BAE}) and relation (\ref{E2}) or by direct diagonalization of the $H^{(2)}$ (\ref{2dt}).
$E^{(2)}_{hom}$ can be obtained by Eqs.(\ref{logBAEt}) (or the reduced BAEs (\ref{BAEt1})) and (\ref{redE2}).

\begin{figure}[!htp]
    \begin{minipage}[t]{0.5\linewidth}
    \centering
    \includegraphics[height=4.5cm,width=8cm]{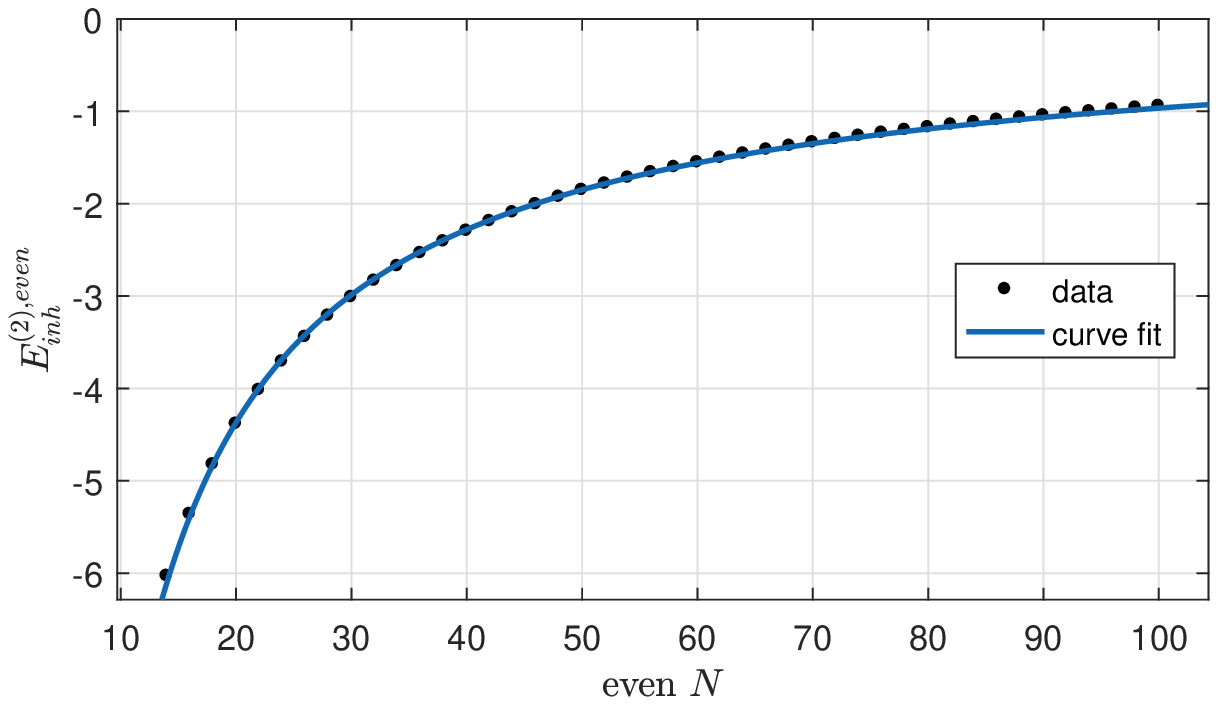}
    \caption*{(a)}
    \end{minipage}
    \begin{minipage}[t]{0.5\linewidth}
    \centering
    \includegraphics[height=4.5cm,width=8cm]{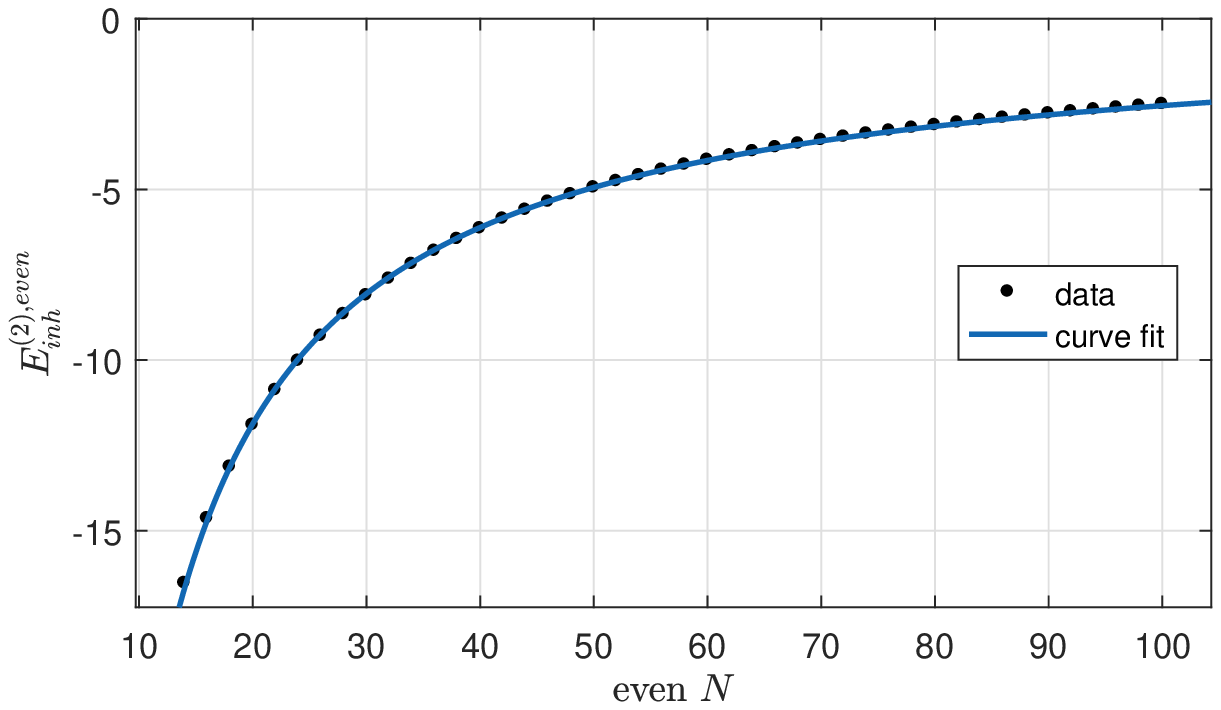}
    \caption*{(b)}
    \end{minipage}
    \caption{The contribution of the inhomogeneous term to the higher conserved charge $ E^{(2),even}_{inh}$ versus the even system-size $N$.
The data can be fitted as $ E^{(2),even}_{inh}(N)=a_8 N^{b_8}$.
Here (a) $\eta=2$, $a_8=-72.46$ and $b_8=-0.9376$; (b) $\eta=3$, $a_8=-207.8$ and $b_8=-0.9558$. Due to the fact $b_8< 0$, the contribution of the inhomogeneous term tends to zero when the $N \rightarrow \infty$.}\label{fig-FC2j-even}
\end{figure}

We consider the $E^{(2)}_{inh}$ corresponding to the ground state of Hamiltonian (\ref{Hamiltoniant}).
In the even $N$ case, we can obtain that $ E^{(2),even}=0$ ($N$ up to 100) for both double degenerate ground states of Hamiltonian (\ref{Hamiltoniant}) by using the DMRG method.
The $E^{(2),even}_{hom}$ can be obtained by Eqs.(\ref{redE2}) and   (\ref{logBAEt}) which the Bethe roots determined by the quantum number (\ref{qnumber-t-g-even}).
From Eq. (\ref{cE2}), the contribution of the inhomogeneous term can be calculated and the results are shown in Fig.\ref{fig-FC2j-even}.
From the fitting, we find that $E^{(2),even}_{inh}$ and $N$ satisfy the power law
\begin{equation}
  E^{(2),even}_{inh}(N)=a_8 N^{b_8}.
\end{equation}
Due to the fact that $b_8<0$, the value of $E^{(2)}_{inh}$ tends to zero when the system-size $N$ tends to infinity, which means that the contribution of the inhomogeneous term can be neglected in the thermodynamic limit.

In the odd $N$ case, we can obtain that $ E^{(2),odd}=0$ ($N$ up to 99) for both double degenerate ground states of Hamiltonian (\ref{Hamiltoniant}) by using the DMRG method.
The $E^{(2),odd}_{hom}$ can be obtained by Eqs.(\ref{redE2}) and   (\ref{logBAEt}) which the Bethe roots determined by the quantum number (\ref{qnumber-t-g-odd}).
It is easy to show that $E^{(2),odd}_{hom}=0$,  which  implies
\begin{equation}
 E^{(2),odd}_{inh}(N)=0.
\end{equation}
Therefore, there is no correction to the higher order conserved charge corresponding to the second order logarithm derivative of the transfer matrix in the odd $N$ case.

%======================================================================================================

\end{document}